\documentclass[10pt, conference, compsocconf]{IEEEtran}

\usepackage {amsmath}
\usepackage {amssymb}
\usepackage{color}
\usepackage {plaatjes}
\usepackage {xspace}
\usepackage {cite}
\usepackage {url}

\definecolor {namecolor} {rgb} {0.9,0.1,0.3}
\definecolor {remarkcolor} {rgb} {0.1,0.7,0.2}

\newtheorem {theorem} {Theorem}
\newtheorem {lemma} {Lemma}
\newtheorem {definition} {Definition}

\newcommand{\routing}{\textsf{ROUTING}\xspace}
\DeclareMathOperator{\diam}{diam}
\DeclareMathOperator{\md}{memdim}
\DeclareMathOperator{\member}{cat}
\DeclareMathOperator{\dist}{sp}
\DeclareMathOperator{\depth}{depth}
\DeclareMathOperator{\height}{height}
\DeclareMathOperator{\ancestors}{ancestors}
\DeclareMathOperator{\leftrm}{left}
\DeclareMathOperator{\rightrm}{right}

\begin {document}

\title{Category-Based Routing in Social Networks:\\ 
    Membership Dimension and
   the Small-World Phenomenon}
   
\author{\IEEEauthorblockN{David Eppstein\IEEEauthorrefmark{1}}\IEEEauthorblockA{\url{eppstein@uci.edu}}
\and \IEEEauthorblockN{Michael T. Goodrich\IEEEauthorrefmark{1}}\IEEEauthorblockA{\url{goodrich@uci.edu}}
\and \IEEEauthorblockN{Maarten L\"offler\IEEEauthorrefmark{1}}\IEEEauthorblockA{\url{mloffler@uci.edu}}
\and \IEEEauthorblockN{Darren Strash\IEEEauthorrefmark{1}}\IEEEauthorblockA{\url{dstrash@uci.edu}}
\and \IEEEauthorblockN{Lowell Trott\IEEEauthorrefmark{1}}\IEEEauthorblockA{\url{ltrott@uci.edu}}
\and 
\and \IEEEauthorblockA{\IEEEauthorrefmark{1}Deptartment of Computer Science, University of California, Irvine, USA}
}


\pagestyle{plain}

\maketitle

\begin{abstract}
A classic experiment by
Milgram shows that individuals can route messages along short paths
in social networks,
given only simple categorical information about recipients (such
as ``he is a prominent lawyer in Boston'' or ``she is a Freshman sociology
major at Harvard'').
That is, these networks have very short paths between pairs of nodes (the so-called \emph{small-world phenomenon});
moreover, participants are able to route messages along these paths
even though each person is only aware of a small part of the 
network topology. Some sociologists conjecture that participants in such scenarios use a \emph{greedy routing} strategy in which they forward messages to acquaintances that have more categories in common with
the recipient than they do, and similar strategies have recently been proposed for routing messages in dynamic ad-hoc networks of mobile devices.
In this paper, we introduce a network property called
\emph{membership dimension}, 
which characterizes the cognitive 
load required to maintain relationships between participants and
categories in a social network.
We show that any connected network has a system of categories that will support greedy routing, but that these categories can be made to have small membership dimension if and only if the underlying network exhibits the small-world phenomenon.
\end{abstract}

\section{Introduction}

In a pioneering experiment in the 1960's, Stanley Milgram and 
colleagues~\cite{milgram67,travers-milgram69,korte-milgram70}
studied message routing in real-world social networks.
296 randomly chosen people in
Nebraska and
Kansas were asked to route a letter to a lawyer in 
Boston by forwarding it to an acquaintance, who would receive the same instructions.
Messages that reached their destinations typically passed between at most six
acquaintances;\footnote{This observation has also led to the 
    concept of ``six degrees of separation''
    between all people on earth
    and the trivia game, ``Six Degrees of Kevin Bacon,'' where players
    take turns trying to link performers
    to the actor Kevin Bacon
    via at most six movie collaborations.} The observation that acquaintance graphs have such short paths has come to be called the 
\emph{small-world phenomenon}~\cite{k-swpa-00,w-ndswp-99}.

Even more surprising than the existence of these short 
paths is that participants are able to 
efficiently route messages using only local information and 
simple facts about targets, such as ethnicity, occupation, name, and location. 



As a way to model the methods used by humans to route such messages, 
sociologists have studied 
the importance of \emph{categories},
that is, various groups to which people belong,
in the small-world phenomenon.
In the early 1970's,
Hunter and Shotland~\cite{hs-tdcsw-74} found that messages routed
between people in the same university category (such as student, faculty,
etc.) had shorter paths than messages routed across
categories.
Killworth and Bernard~\cite{Killworth-Reverse} 
performed experiments in the late 1970's that they called 
\emph{reverse small-world experiments} in which each participant was presented with a list of
messages for hundreds of targets, identified by the categories of 
town, occupation, ethnic background, and gender,
and asked to whom they would
send each of these messages.
The study concluded that the choices people
make in selecting routes are overwhelmingly categorical in nature.
In the late 1980's, Bernard {\it et al.}~\cite{bkems-ssrcc-88}
extended this work to identify which of twenty categories are most
important for message routing to people from various cultures.
More recently,
Watts {\it et al.}~\cite{Watts02identityand} 
present a hierarchical model for categorical organization in social 
networks for the sake of message routing. They propose groups
as the leaves of rooted trees, with internal nodes defining
groups-of-groups, and so on. They define an ultrametric on sets of
such overlapping hierarchies and conjecture that people use the
minimum distance in one of their trees to make message routing 
decisions.
That is, they argue that
individuals can understand their ``social distance" to a target
as the minimum distance between them and the target in one of their
categories.
Such a determination requires some global knowledge about
the structures of the various group hierarchies.

Although this previous work shows the importance of categories and of hierarchies of categories in explaining the small world phenomenon, it does not explain where the categories come from or what properties they need to have in order to allow greedy routing to work.  
Hence, this prior work leaves open the following questions:
\begin{itemize}
\item
Which social networks support systems of categories that allow participants to route messages using the simple greedy rule of sending
a message to an acquaintance who has more categories in common with
the target? 
\item
How complicated a system of categories is needed for this purpose, and what properties of the underlying network can be used to characterize the complexity of the category system?
\end{itemize}



Our goal in this paper, therefore, is to address these questions
by studying the existence
of mathematical and algorithmic 
frameworks that demonstrate the feasibility of local, greedy, category-based
routing in social networks.

\subsection{Our Results}

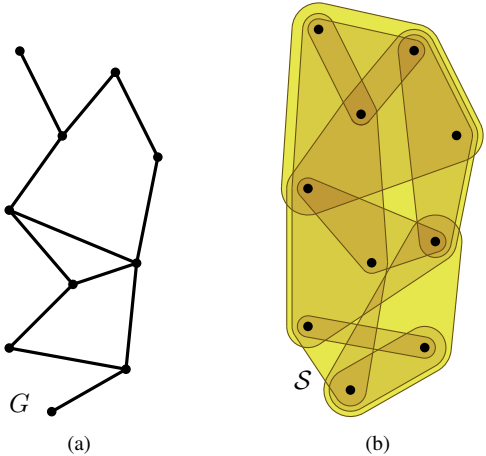
\tweeplaatjes {ex-graph} {ex-groups} {A set of elements $U$ (drawn arbitrarily as points in the plane). (a) The graph $G$ on $U$. (b) The categories $\mathcal S$ on $U$. In this example, the membership dimension is $4$, because no element is contained in more than $4$ groups.}

Inspired by the work of
Watts {\it et al.}~\cite{Watts02identityand},
we view a social network as an undirected graph $G=(U,E)$, whose
vertices represent people and whose edges 
represent relationships, taken together with a collection,
${\mathcal S} \subset 2^{U}$, of categories defined on the vertices
in $G$. 
Figure~\ref {fig:ex-graph+ex-groups} shows an example.
In addition, 
given a network $G=(U,E)$ and category system ${\mathcal S}$,
we define the \emph{membership dimension}
of $\mathcal S$ to be 
\[
\max_{u\in U} |\{ C\in {\mathcal S}\colon\, u\in C\}|,
\]
that is,
the maximum number of groups to which any one person in the network 
belongs.  The membership dimension characterizes the \emph{cognitive load} of performing routing tasks in the given system of 
categories---if the membership dimension is small, each actor in the network only needs to know a proportionately small amount of information about his or her own categories, 
his or her neighbors' categories,
and the categories of each message's eventual destination.
Thus, we would expect real-world social networks to have small
membership dimension.

In this paper, we provide a constructive proof that a category system can support greedy routing. Our results are not intended to model the actual formation of social categories, and we take no position on whether categories are formed from the network, the network is formed from categories, or both form together. Rather, our intention is to show the close relation between two natural parameters of a social network, its path length and its membership dimension.
In particular:
\begin{itemize}
\item
We show that the membership dimension 
of $(G,{\mathcal S})$
must be at least the diameter of $G$, $\diam(G)$, 
for a local, greedy, category-based routing strategy to work.
\item
We show that every connected graph $G=(U,E)$, 
has a collection $\mathcal S$ of categories
such that local, greedy, category-based routing always works,
with membership dimension $O((\diam(G)+\log |U|)^2)$.
\end{itemize}
Since Milgram's work~\cite{milgram67,travers-milgram69,korte-milgram70},
social scientists have believed that real-world social
networks have diameters bounded by  constants or  slowly growing functions of the network size.
Under a weak form of this assumption, that the diameter is $O(\log |U|)$,
our results provide a natural model for how participants in a social
network could efficiently route messages 
using a local, greedy, category-based routing strategy
while remembering an amount of
information that is only polylogarithmic in the size of the network.

%

\subsection{Previous Related Work}

Geometric greedy routing~\cite{Couto01locationproxies,Kranakis99compassrouting}
uses geographic location rather than categorical data to route
messages.  
In this method, vertices have coordinates 
in a geometric metric space and messages are routed to any neighbor that is closer to the target's coordinates.
Greedy routing may not succeed in certain geometric networks, so a number of
techniques have been developed to assist such greedy routing schemes
when they fail~\cite{bose-GaurDel-01,Karp:2000:GGP:345910.345953,Kuhn:2003:GAR:872035.872044}.
Introduced by Rao et al.~\cite{Rao:2003:GRW:938985.938996},
virtual coordinates can overcome the shortcomings of 
real-world coordinates and allow simple greedy forwarding to function 
without the assistance of fallback algorithms. This approach
has been explored by other 
researchers~\cite{Papadimitriou:2005:CRG:1121826.1121828,leighmoit-someresul-10,Angelini-GDTriangulations-09,Klein-Hyp-07},
who study various network properties that allow for greedy routing to
succeed.
Several researchers also study the existence of 
\emph{succinct} greedy-routing strategies~\cite{Muh-dis-07,Maymounkov06greedyembeddings,Eppstein:2009:SGG:1506879.1506884,Goodrich-SuccEuclid-09},
where the number of bits needed to represent the coordinates of each vertex is polylogarithmic in the
size of the network; this notion of succinctness for geometric greedy routing is closely analogous to our definition of the membership dimension for categorical greedy routing.

Recent work by Mei {\it et al.}~\cite{Mei-Cat-11},
studies category-based greedy routing as a heuristic
for performing routing in dynamic delay-tolerant networks. Mei {\it et al.} assume that
the network nodes have been organized into pre-defined categories based on
the users' interests. 
Experiments suggest that using these categories for greedy routing is superior to
routing heuristics based on location or simple random choices.
One can interpret the categorical greedy routing techniques of Mei {\it et al.} and of this paper
as being geometric routing schemes using virtual coordinates, where the coordinates represent category memberships. In this interpretation, the membership dimension of an embedding corresponds to the number of nonzero coordinates of each node, and our results
show that such greedy routing schemes can be done succinctly in
graphs with small diameter.

Similarly to the work of this paper,
Kleinberg~\cite{k-swpa-00} studies the 
small-world phenomenon from an algorithmic perspective.
However, his approach is orthogonal to ours: He focuses on location rather than categorical information as the critical factor for
the ability to find short routes efficiently, and constructs
a random network based on that information, whereas
our approach takes the network as a given and studies the kinds of
categorical structures needed to support category-based greedy
routing.

In addition, it is worth noting that small world networks exhibit scale-free properties.




\section {Routing based on Categorical Information}

In this section, 
we introduce a mathematical model of categorical greedy routing,  and provide basic definitions and properties that guarantee the success of this strategy.
  \subsection {Basic definitions}
  
    Abstracting away the social context, let $U$ be the universe of
    $n$ people defining the potential sources, targets, and intermediates for
    message routes, and
    let $G=(U,E)$ be an undirected graph whose $m$ edges represent 
    pairs of people who can communicate. 
    For any two elements $s, t \in U$, let $\dist(s,t)$ be the length of the shortest path in~$G$ from $s$ to~$t$. The diameter $\diam(G)= \max_{s,t \in U} \dist(s,t)$ is the maximum length of any shortest path. 
   For $s \in U$, 
     define the \emph {neighborhood} of~$s$
    to be the set of neighbors $N(s) = \{u \in U \mid \{s,u\} \in E\}$ of $s$ in $G$.
    
    Now let ${\mathcal S} \subset 2^{U}$ be a set of subsets of $U$, which represent the abstract categories that elements of $U$  belong to.
    For a given $u \in U$, we define $\member(u) \subset \mathcal S$ to be the set of groups to which $u$ belongs:
    $\member(u) = \{C \in \mathcal S \mid u \in C\}$.


    \begin {definition}[membership dimension]
      The \emph {membership dimension} of $\mathcal S$ is the maximum number of elements of $\mathcal S$ that any element of $U$ is contained in, that is,
      \[
      \md(\mathcal S) = \max_{u\in U} |\member (u)|.
      \]
    \end {definition}

    As discussed, there is evidence that in real world social networks and group structures $(G,\mathcal S)$, both $\diam (G)$ and $\md(\mathcal S)$ are significantly smaller than $|U|$.


  \subsection {The routing strategy}

    We now describe a simple greedy category-based strategy to route a message from one node to another. We clarify the distance function immediately following the rule definition.
  
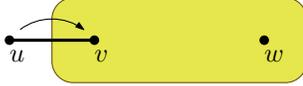
    \eenplaatje {routing} {Illustration of the routing rule.}

    \begin {definition}[greedy routing rule]
      If a node $u$ receives a message $M$ intended for a destination $w \neq u$, then $u$ should forward $M$ to a neighbor $v \in N(u)$ that is closer to $w$ than $u$ is, that is, for which $d(v,w) < d(u,w)$.
    \end {definition}


    
The category-based distance function used by this rule is  $d(s,t) = |\member(t) \setminus \member(s)|$, which measures the number of categories of the target that the current node does \emph {not} share.\footnote{Note that $d$ is not a metric, since it is not necessarily symmetric.} This number decreases as the number of shared groups of $\mathcal S$ between the current node and the target increases. We refer to the greedy routing strategy that uses this distance function as \routing(see Figure~\ref {fig:routing}).

    For category systems with low membership dimension, this strategy is easy to evaluate using only local knowledge about the categories of each neighbor of the current node and the categories of the target node.


  \subsection {Successful routing}
  
    We now investigate  conditions under which \routing can successfully route messages between all pairs of nodes in a network. We identify several properties of a graph $G$ and associated group structure $\mathcal S$ 
    that directly influence the feasibility of  routing.
    For routing to succeed, $G$ must be connected.
    It seems natural to consider a stronger property:

    \begin {definition}[internally connected]
      $(G, \mathcal S)$ is  
      \emph{internally connected}
      if for each $C \in \mathcal S$, $G$ restricted to $C$ is connected.
    \end {definition}

    Figure~\ref {fig:ic-not-sh} shows an example of an internally connected pair $(G, \mathcal S)$.
    This is a very natural property for sociological groups to exhibit. 
    People belonging to the same group will have greater cohesiveness, 
    and if a group  is not internally connected
    then it may be redefined to be the 
    set of groups defined by its connected components.

    \tweeplaatjes {ic-not-sh} {sh-not-ic} 
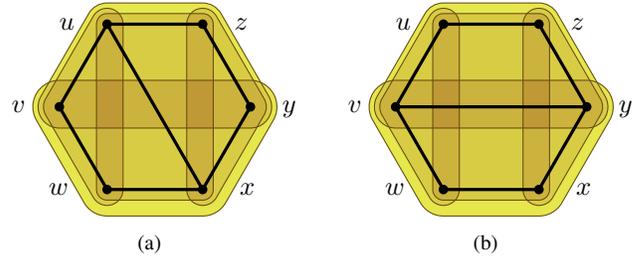
    {Two networks with the same elements and categories.  (a) An example that is internally connected, but not shattered:  no category contains $y$ and a neighbor of $v$ but not  $v$ itself. (b) An example that is shattered, but not internally connected: the induced graph of category $\{u,w,x,z\}$ is not connected.}
  
    \begin {definition}[shattered]
      A pair $(G, \mathcal S)$ is \emph {shattered} if,
      for all $s, t \in U$, $s \neq t$, there is a neighbor $u \in N(s)$
      and a set $C \in \mathcal S$ such that $C$ contains $u$ and $t$, but not $s$.
    \end {definition}

    Figure~\ref {fig:sh-not-ic} shows an example of a shattered pair.
    Note that in this definition, $u$ and $t$ could be the same node.
    This property falls out naturally from the instructions given in the
    real-world routing experiments of Milgram and others. 
    In order for someone to advance a letter toward a target, 
    there must be an acquaintance that shares additional interests 
    with the target. 
    Indeed, we now show that the shattered property is 
    necessary for \routing to work.

\begin{lemma} \label {lem:shattered-necessary}
If $(G, \mathcal{S})$ is not shattered,  \routing fails.
\end{lemma}

\begin{proof}
Since $(G, \mathcal{S})$ is not shattered, there exists a pair of vertices $s$ and $t$, where
$s$'s neighbors are not in sets with $t$ that do not contain $s$.
Therefore, $s$'s neighbors
cannot share strictly more sets with $t$ as $s$ does, and \routing will fail
to route from $s$ to $t$.
\end{proof}

If $G$ is a tree, then these two properties together are sufficient for the routing strategy to always work:

  \begin {lemma}
\label{lemma:treeroutingworks}
    If $G$ is a tree, and $(G,\mathcal{S})$ is internally connected and shattered, then \routing is guaranteed to work.
  \end {lemma}

\begin{proof}
Let $s$ and $t$ be vertices in $G$. Since $G$ is a tree,
there is one simple path from $s$ to $t$. 
Let $(u,v)$ be an edge on the path from $s$ to $t$. 
First, we claim that every set in $\mathcal S$ that contains both $u$ and
$t$ also contains $v$. This follows from $(G, \mathcal S)$ being internally 
connected: any set $C \in \mathcal S$ with $u, t \in C$ must also contain $v$, 
since $v$ is on the only path between $u$ and $t$. 
Therefore, $v$ is contained in at least as many sets in $\mathcal S$
with $t$ as $u$ is. 
However, by the shattered property, $v$ is in a set in $\mathcal S$ 
with $t$ that does not contain $u$.
Therefore $v$ is in strictly more sets with $t$ than $u$ is. 
This property holds for every simple path; hence,
\routing always works.
\end{proof}

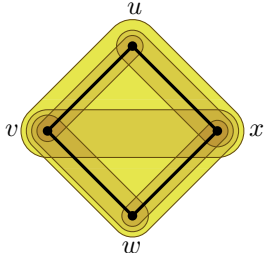
\eenplaatje {counter} {\routing  does not work in this graph, even though it is internally connected and shattered.   Routing from $v$ to $x$ fails: $v$, $u$, and $w$ are all at distance 2 from $x$, so $v$ has no neighbor that is closer than it to $x$.}
 
Although sufficient for routing in trees,
the internally connected and shattered properties are not sufficient 
for \routing to work on arbitrary connected graphs. 
Figure~\ref{fig:counter} shows a 
counter-example---\routing
is unable to route a message from the leftmost to the rightmost node, 
since there is no neighbor whose distance to the target is smaller.

\section {Existence of Categories}

In this section, we consider the following question: Is it possible to construct the family $\mathcal{S}$ so that
\routing always works and $\mathcal{S}$ has low membership dimension? 

We show that such a construction is always
possible if we are given a connected graph as input. We also show that it is impossible to construct
an $\mathcal{S}$ such that \routing will work if the graph is not known in advance. 

\subsection {Constructing $\mathcal S$ given $G$}

    Given a connected graph $G=(U,E)$ as input, we would like to construct a family $\mathcal{S}\subset 2^{U}$ so that
      \routing works, and the membership dimension of $S$ is small. 
      We concentrate foremost on constructions
    of category collections
    that are internally connected and shattered, 
    because of the social significance of these
    properties. 
    Nevertheless, even without these properties, we have the following lower bound.

    \begin {lemma}
      Let $G$ and $\mathcal S$ be a graph and a category system, respectively, such that  \routing works for $G$ and $\mathcal S$. Then $\md(\mathcal S) \geq \diam (G)$.
    \end {lemma}

    \begin {proof}
      By definition of the diameter, there are two vertices $s, t \in U$
      such that $sp(s, t) = \diam(G)$.
      Let $P$ be the path that \routing follows from $s$ to $t$,
      and note that the length of $P$ must be 
      at least $\diam (G)$.
      An edge $(u,v)$ can only be on $P$ if $d(v,t) < d(u,t)$.
      Since $d(\cdot,\cdot)$ can only take integer values, $d(u,t) \geq d(v,t) + 1$.
      Therefore, $d(s,t) \geq |P|$.
      By definition, $d(s,t) = |\member(t) \setminus \member(s)|$, and $\md(\mathcal S)$ is the maximum of $\member (\cdot)$ over all elements; hence
      $
        \md(\mathcal S) \geq |\member(t)| \geq |\member(t) \setminus \member(s)|
        = d(s,t) \geq |P| \geq \diam(G),
      $
      as claimed.
    \end {proof}

    For paths, this bound is tight:

    \begin {lemma}
    \label{lem:pathsic}
      If $G$ is a path, then there exists an $\mathcal S$ s.t. $(G, \mathcal S)$ is shattered and internally connected with $\md(\mathcal S) = \diam(G)$.
    \end {lemma}

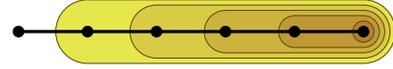
    \eenplaatje {paths} {The sets $B_v$ for each vertex $v$ in the path. The sets $A_v$ are constructed symmetrically. }

    \begin {proof} Arbitrarily pick one of the two end vertices of $G$ and 
    let us refer to the vertices in $G$ by their distance, $0$ to $n-1$, 
    from this vertex.
    For each vertex $i$, form two sets $A_i$ and $B_i$, 
    where $A_i=\{0,\ldots,i-1\}$ and $B_i=\{i+1,\ldots,n-1\}$, 
    and let $\mathcal{S} = \bigcup_{v\in U} \{A_v, B_v\}$.
    Figure~\ref{fig:paths} illustrates this construction.
    Each set in $\mathcal{S}$ consists of a path of vertices and therefore $\mathcal{S}$ is internally connected.
    $\mathcal{S}$ is also shattered, since for all $s$ and $t$, $s$ has a neighbor that shares either $A_s$ or $B_s$ with $t$, but $s$ is not in these sets.
    Considering $\md(\mathcal{S})$, note that each vertex $i$ 
    is contained in sets $A_j$ for $0\leq j<i$ and $B_k$ for $k<i\leq n-1$.
    Therefore, each vertex is in exactly $n-1$ sets, which is $\diam(G)$. 
    \end {proof}

It follows from Lemmas~\ref{lemma:treeroutingworks} and~\ref{lem:pathsic} that, if $G$ is a path,
one can construct $\mathcal{S}$ with 
$\md(\mathcal{S})=\diam(G)$, so that \routing works in $G$.

There are
other graphs for which it is relatively easy 
to set up a category set that is 
shattered and internally connected in a way that supports the
\routing algorithm.
For example, in a tree of height~$1$ (i.e., a star graph),
with root $r$, we could create for each leaf of the tree two categories, one containing the leaf itself and one containing both the leaf and the root.
Every path in this tree supports  \routing.
However, the
membership dimension of this category system is high, since the root
belongs to a linear number of categories. So even in this simple
example, supporting  \routing and achieving 
low membership dimension is a challenge.
Moreover, this challenge becomes even more difficult already for a
tree of height~$2$, since navigating from any leaf, $x$, to another leaf,
$y$, requires that the parent of $x$ belong to more categories with 
$y$ than $x$---and this must be true for \emph{every} other leaf,
$y$.
Thus,
it is perhaps somewhat surprising that we can construct a set of
categories, $\mathcal S$, for an arbitrary binary tree that causes
this network to be shattered and
internally connected (so the \routing strategy works, by
Lemma~\ref{lemma:treeroutingworks}) and such that $\mathcal S$
has small membership dimension.

    \begin{lemma} \label{lemma:binarytreememdim}
        If $G$ is a binary tree, 
  then there exists an $\mathcal S$ s.t. $(G, \mathcal S)$ is shattered and internally connected 
  with $\md(\mathcal S) = O(\diam^2(G))$.
    \end{lemma}

    \begin{proof} 
    We show how to construct $\mathcal{S}$ from $G$.
Arbitrarily pick a vertex $r\in U$ of degree at most $2$ and root the binary tree at $r$,
so each vertex $v$ has left and right children,
$\leftrm(v)$ and $\rightrm(v)$,
and let $\height(v)$ be the length of the longest simple path from $v$ to any descendant of $v$.
For each vertex $v$, we create a set $S_v$, containing $v$'s descendants (which includes $v$).
We further construct two families, $L_v$ and $R_v$,
using helper sets $L_{v,i}$ and $R_{v,i}$.
Let $L_{v,i}$ (resp., $R_{v_i}$) consist of $v$, 
the vertices in $v$'s left (right) subtree down to depth $i$,
and all vertices in $v$'s right (left) subtree.
Then define
\[
L_v = \{L_{v,i}\mid \depth(v)\leq i \leq \depth(v)+\height(\leftrm(v))\}.
\]
Figure~\ref {fig:bintree} illustrates this.
The family $R_v$ is defined symmetrically. Our $\mathcal{S}$ is then 
defined as
\[
{\mathcal S}\,  =\, \bigcup_{v\in U}{\{S_v\}\cup L_v \cup R_v}.
\]
Each set in $\mathcal{S}$ is a connected subgraph of $G$, so $\mathcal{S}$ is internally connected.
As the following argument shows, $\mathcal{S}$ is shattered:
If $s$ is an ancestor of $t$, then $s$'s child $u$ on the path to $t$ is in set $S_u$ which contains $u$ and $t$ but not $s$.
Otherwise, let $v$ be the lowest common ancestor of $s$ and $t$, and assume without loss of generality that $s$ in $v$'s left subtree; then $L_{v,\depth(s)-1}$ contains $s$'s parent and $t$ but not~$s$.

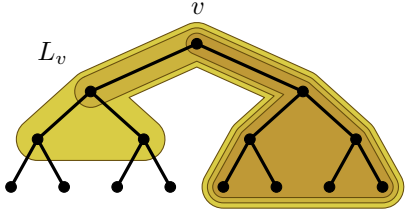
    \eenplaatje {bintree} {The collection of sets $L_v$ for an example subtree at $v$.}

We now analyze the membership dimension of this construction.
Let $v$ be a vertex, and let $\ancestors(v)$ be the set of $v$'s ancestors.
For $u\in \ancestors(v)$, $v\in S_u$, and $v$ belongs to $O(\height(u))$ sets of $L_u$ and $R_u$.
Then $v$ belongs to $O\left(\sum_{u\in \ancestors(v)}\height(u)\right)$ sets,
which is $O(\diam^2(G))$ for any $v$.
    \end{proof}

We now extend this result to arbitrary trees by applying \emph{weight-balanced binary trees}\cite{knuth71,bent85}.

\begin{definition}[weight balanced binary tree]
A weight balanced binary tree is a binary tree
that stores weighted items in its leaves. If item $i$
has weight $w_i$, and all items have a combined weight of $W$ then item 
$i$ is stored at depth $O(\log {(W/w_i)})$. 
\end{definition}

\begin{lemma}
\label{lemma:embedtree}
Let $T$ be an $n$-node rooted tree with height $h$. We can embed $T$ into a binary tree such that the ancestor--descendant relationship is preserved, and the resulting tree has height $O(h + \log n)$.
\end{lemma}

\begin{proof}
Let $n_u$ be the number of descendants of vertex $u$ in $T$. For each vertex $u$ in $T$ that has more than two children, we expand the subtree consisting of $u$ and $u$'s children into a binary tree as follows. Construct a weight balanced binary tree $B$ on the children of $u$, where the weight of a child $v$ is $n_v$. We let $u$ be the root of $B$. Each child $v$ of $u$ in the original tree is then a leaf at depth $\log (n_u/n_v)$ in $B$. 
Performing this construction for each vertex $u$ in the tree expands $T$ into a binary tree with the ancestor--descendant relationship preserved from $T$.

Furthermore, each path from root to leaf in $T$ is only expanded 
by $\log (n)$ nodes, which we can see as follows.
Each parent-to-child edge $(u,v)$ in $T$ is replaced by a path of length 
$O(\log (n_u/n_v))$.
Therefore for each path $P$ from root $r$ to leaf $l$ in $T$, our construction expands $P$ by length $O(\sum_{(u,v)\in P}\log (n_u/n_v))$, 
which is a sum telescoping to $O(\log (n_r/n_l)) = O(\log n)$. 
Therefore, the height of the new binary tree is $O(h + \log n)$.
\end{proof}

Combining this lemma with Lemma~\ref{lemma:treeroutingworks}, we get the following theorem.

\begin{theorem}
\label{theorem:routetree}
Given a tree $T$, it is possible to construct a family $\mathcal{S}$ of subsets such that \routing works 
for $T$ and $\md(\mathcal{S}) = O((\diam(T) + \log n)^2)$.
\end{theorem}

\begin{proof}
Arbitrarily root $T$ and embed $T$ in a binary tree $B$ using the method in 
Lemma~\ref{lemma:embedtree}. Then $B$ has height $O(\diam(T) + \log n)$, and diameter $\diam(B) = O(\diam(T) + \log n)$. 
Applying the construction from Lemma~\ref{lemma:binarytreememdim}
to $B$ gives us a family $\mathcal{S}_B$
with $\md(\mathcal{S}_B)=O((\diam(T) + \log n)^2)$. 
We then construct a family $\mathcal{S}_T$, 
by removing vertices that are in $B$ but not $T$ from the sets in $\mathcal{S}_B$.
By construction, $(T,\mathcal{S}_T)$ is shattered and internally connected, and $\md(\mathcal{S}_T)\leq \md(\mathcal{S}_B)= O((\diam(T) + \log n)^2)$. By Lemma~\ref{lemma:treeroutingworks}, \routing works on $T$ with category
sets from $\mathcal{S}_T$.
\end{proof}

We can further extend this theorem to arbitrary connected graphs,
which is the main upper bound result of this paper.

\begin{theorem}
\label{theorem:memdimG}
      If $G$ is connected, there exists $\mathcal S$ s.t. \routing works and $\md(\mathcal S) = O ((\diam(G)+\log(n))^2)$.
\end{theorem}

\begin{proof}
Compute a low-diameter spanning tree $T$ of $G$. This step can easily be done using breadth-first search, producing a tree with diameter at most $2\diam(G)$. We then use the construction from Theorem~\ref{theorem:routetree} on $T$.
For greedy routing to work in a graph $G$,
note that
it is sufficient to show that it works in a spanning tree of $G$.
Therefore, since \routing works in $T$, \routing also works in $G$. 
\end{proof}

\section{Conclusion and Open Problems}

We have presented a construction of groups $S$ on a connected graph $G$
that allows a simple greedy routing algorithm, utilizing a
notion of distance on group membership, to guarantee delivery between nodes in 
$G$.
Such a construction will have membership dimension $O ((\diam(G)+ \log n)^2)$,
demonstrating a small cognitive load for the members
of $G$.

There are several directions for future work.
For example,
while we have shown that the membership dimension must 
be minimally the diameter of $G$, it remains to be shown 
if the membership dimension must be the square of the diameter plus a
logarithmic factor for
arbitrary graphs.
We conjecture that the square term is not strictly needed in the
membership dimension in order for \routing to work.
Our group construction is performed for a general graph by selecting a low diameter spanning tree and using the presented tree construction,
so it may be possible that there is a group construction that 
has lower membership dimension and more efficient routing if 
it is constructed directly in $G$.

In this paper all categories are given equal
weight with respect to routing tasks and that participants use
a simple greedy routing algorithm based solely on increasing the
number of categories in common with the target.
Future work could include study of a
category-based routing strategy that allows participants to weight
various categories higher than others, as
in the work of Bernard {\it et al.}~\cite{bkems-ssrcc-88}.




  \bibliographystyle {abbrv}
  \bibliography {refs,greedy}

\end {document}